%% file: main_acm.tex
\documentclass[format=sigconf, review=false, anonymous=false, dvipsnames]{lib-acm/acmart}

\copyrightyear{2026}
\acmYear{2026}
\setcopyright{cc}
\setcctype{by}
\acmConference[GI 2026]{Graphics Interface 2026}{June 09--12, 2026}{Waterloo, ON, Canada}
\acmBooktitle{Graphics Interface 2026 (GI 2026), June 09--12, 2026, Waterloo, ON, Canada}
\acmDOI{10.1145/3831423.3831427}
\acmISBN{979-8-4007-2666-8/2026/06}

\usepackage{booktabs} %

\input{_setup.tex}

\begin{document}

\tolerance=400 

\title[Interaction Techniques to Encourage Longer Prompts]{Interaction Techniques that Encourage Longer Prompts Can Improve Psychological Ownership when Writing with AI}

\author{Nikhita Joshi}
\orcid{0000-0001-9493-7926}
\affiliation{%
  \institution{Cheriton School of Computer Science\\University of Waterloo}
  \city{Waterloo, Ontario}
  \country{Canada}
}
\additionalaffiliation{%
  \institution{LISN, Universit\'{e} Paris-Saclay, CNRS, Inria}
\city{Orsay}
  \country{France}
}
\email{nvjoshi@uwaterloo.ca}

\author{Daniel Vogel}
\orcid{0000-0001-7620-0541}
\affiliation{%
  \institution{Cheriton School of Computer Science\\University of Waterloo}
    \city{Waterloo, Ontario}
  \country{Canada}
}
\email{dvogel@uwaterloo.ca}

\renewcommand{\shortauthors}{Joshi and Vogel}

\begin{abstract}
\input{_abstract}
\end{abstract}



\begin{teaserfigure}
\centering
  \includegraphics[width=\textwidth]{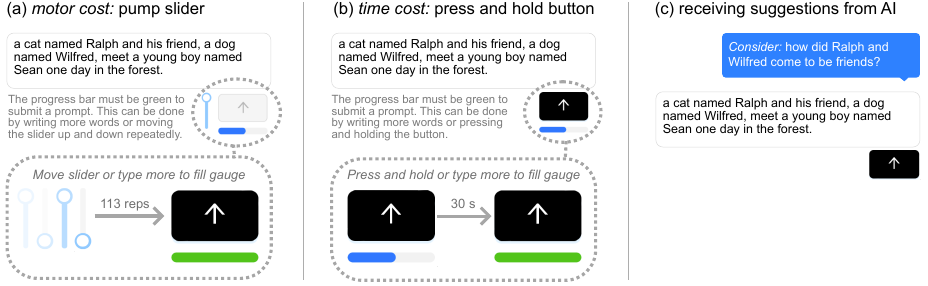}
  \caption{Augmented prompt entry techniques that (a) increase the motor cost of submitting a prompt by making the user `pump' a gauge by moving a slider; (b) increase the temporal cost by requiring users to press and hold the button to fill a gauge; and (c) provide AI-generated suggestions for how the user's prompt can be expanded.}
  \label{fig:teaser}
\end{teaserfigure}

\maketitle

\input{_body.tex}

\bibliographystyle{lib-acm/ACM-Reference-Format}
\bibliography{main_acm}

\appendix
\input{8-appendix}

\end{document}

%% file: _setup.tex
\usepackage{exii-macros}

\usepackage{booktabs}
\usepackage{multirow}

\makeatletter
\g@addto@macro\normalsize{%
  \setlength\abovedisplayshortskip{-9pt}
  \setlength\belowdisplayshortskip{3pt}
}
\makeatother

%% file: _abstract.tex
Writing longer prompts for an AI assistant to generate a story increases psychological ownership, a user's feeling that the writing belongs to them. To encourage users to write longer prompts, we evaluated two interaction techniques that modify the prompt entry interface of chat-based generative AI assistants: pressing and holding the prompt submission button, and continuously moving a slider up and down when submitting a short prompt. A within-subjects experiment investigated the effects of such techniques on prompt length and psychological ownership, and results showed that these techniques increased prompt length and led to higher psychological ownership than baseline techniques. A second experiment further augmented these techniques by showing AI-generated suggestions for how the prompts could be expanded. This further increased prompt length, but did not lead to improvements in psychological ownership. Our results show that simple interface modifications like these can elicit more writing from users and improve psychological ownership.

%% file: _body.tex
\input{1-introduction}

\input{2-background}

\input{3-technique}

\input{4.1-exp-protocol}

\input{4.2-exp-results}

\input{5-exp2}

\input{5.1-comparing}

\input{6-discussion}

\input{7-conclusion}

%% file: 1-introduction.tex
\section{Introduction}
Psychological ownership is a concept in psychology that describes possessive feelings people have towards anything, for example, ideas, places, objects, and even creations, regardless of legal ownership \cite{Pierce2001PsychologicalOwnership, Pierce2003PsychologicalOwnership}. Writers often feel psychological ownership for written text \cite{nicholes2017measuring, Caspi2011Collaboration, Blau2009GoogleDocs, Halfaker2009WikipediaOwnership, ThomSantelli2009TerritorialWikipedia}, likely because writing requires an investment of time, energy, and sense of self, all of which are important for psychological ownership \cite{Pierce2001PsychologicalOwnership, Pierce2003PsychologicalOwnership, Wang2006PsychologicalOwnershipDigital}. Losing too much psychological ownership can happen when writing collaboratively \cite{Caspi2011Collaboration, Blau2009GoogleDocs, Halfaker2009WikipediaOwnership, ThomSantelli2009TerritorialWikipedia}, which can lead to negative outcomes like territorial behaviours \cite{LarsenLedet2019Territorial, ThomSantelli2009TerritorialWikipedia} and can even deter future collaborations \cite{Caspi2011Collaboration}.

More recently, writers have been `collaborating' with generative AI tools like ChatGPT. Research has shown that psychological ownership is also important for human-AI collaborative writing systems \cite{lee2024design, shen2023parachute}, leading to increased interest within the HCI community in how using these tools affects psychological ownership (e.g., \cite{li2024valueAIOwnership, Lehmann2022SuggestionList, Draxler2024GhostwriterOwnershipStudy}).
Notably, Joshi and Vogel \cite{Joshi2025Ownership} explored the role of \emph{prompt length} on psychological ownership when writing short stories, and found that writing longer prompts increased psychological ownership. Participants were encouraged to provide more details about the story plot; however, this was primarily achieved through a strict word minimum on the prompt length. In practice, users desire more flexibility. One way to encourage longer prompts without preventing shorter prompts is to create interaction techniques that purposely introduce \emph{friction} when users write short prompts by introducing \emph{costs}. As an example, they show that requiring the user to press and hold the prompt submission button, a time cost, can encourage users to write longer prompts, but they did not evaluate the effects of this technique on psychological ownership.

There are many other ways to introduce friction into the prompt submission interface to encourage users to write longer prompts for increased psychological ownership. Inspired by their technique, we propose augmenting the prompt entry interface so users have to fill a `gauge' by writing more words, or by repeatedly moving a slider up and down (Figure \ref{fig:teaser}). Much like Joshi and Vogel's time delay technique, this provides users with a choice: they can either write a longer prompt, or exert physical effort to submit a shorter one. We evaluated the effect of both augmented prompt entry interfaces on prompt length and psychological ownership. Twenty nine participants wrote prompts to generate short stories for a generative AI writing assistant and answered subjective questions about psychological ownership and task workload. When compared to a baseline interface with no augmentations, participants felt more mental demand, physical demand, effort, and frustration, but wrote significantly longer prompts and felt more psychological ownership toward the resulting stories. To encourage participants to generate more ideas \cite{TimedSuggestions2015} and write even longer prompts, we further augmented these techniques to display AI-generated suggestions every 8 seconds, triggered after one second of inactivity. A follow up study with 30 participants showed that this led to significantly longer prompts for the augmented prompt entry conditions, but did not further improve psychological ownership.
Our work contributes:
\begin{itemize}
    \item a new interaction technique for increasing prompt length that increases the motor cost of submitting a short prompt;
    \item empirical evidence that shows this technique improves psychological ownership and confirms the effectiveness of Joshi and Vogel's technique;
    \item empirical evidence that shows that receiving AI-generated suggestions increases prompt length, but does not improve psychological ownership.
\end{itemize}

%% file: 2-background.tex
\section{Background and Related Work}
Our work relates to previous work examining psychological ownership when writing with AI, and previous work that introduces friction into user interfaces.

\subsection{Psychological Ownership when Writing with AI}
Given the increase of writing tools with AI capabilities, researchers have been advocating for AI writing systems that are designed \cite{lee2024design} and evaluated \cite{shen2023parachute} in a way that prioritizes psychological ownership. Some prior work has explored the effects of the role of AI writing assistants on psychological ownership. Biermann et al. \cite{Biermann2022OwnershipLLM} presented writers with mock-up user interfaces that utilized AI assistance in different ways when writing stories (e.g., using AI to generate character dialog, or to elaborate on the scene's setting), and found that many designs were not well-received as they reduced psychological ownership. Draxler et al. \cite{Draxler2024GhostwriterOwnershipStudy} explored using an AI assistant for writing postcards. They found that when the AI assistant wrote the entire message, participants did not feel much psychological ownership, yet did not attribute authorship to AI (i.e., an AI Ghostwriter Effect). 
Li et al. \cite{li2024valueAIOwnership} showed that when AI acted as the writer, psychological ownership was reduced for participants when compared to the AI acting as an editor. However, they were willing to accept lower remuneration in exchange for receiving writing assistance from the AI assistant, signifying its high value.

The type and quantity of AI suggestions during the writing process can affect psychological ownership. Lehmann et al. \cite{Lehmann2022SuggestionList} showed that manually integrating AI-generated suggestions into text led to higher feelings of psychological ownership than those that were automatically integrated. Dhillon et al. \cite{dhillon2024ScaffoldingOwnership} showed that when writers received suggestions, psychological ownership decreased, yet this effect was more pronounced for longer suggestions (paragraphs) than shorter ones (sentences). Lee et al. \cite{Lee2022CoAuthor} speculate that receiving more suggestions from AI reduces how much the human writes, leading to lower psychological ownership.

Joshi and Vogel \cite{Joshi2025Ownership} were the first to provide empirical evidence that the ratio of human-to-AI-generated text can influence psychological ownership. Participants wrote prompts of pre-defined lengths (ranging from 3 to 150 words), which were enforced through word minimums and maximums to generate short stories. As participants wrote longer prompts, they felt more psychological ownership towards the resulting stories. This was likely because participants included more details about the story in their prompts, which required additional thought and effort.

\subsection{Friction in User Interfaces}
Though it may feel counterintuitive to purposely make the act of submitting a prompt more difficult, prior work shows that this can be beneficial. Notably, friction in design can encourage users to move from fast, automatic, and intuitive ``System 1'' thinking to slow, deliberate, and reflective ``System 2'' thinking that is more conducive to increased attention and analytical thinking \cite{kahneman2011thinking, Cox2016DesignFrictions, Caraban2019Nudging, BuccincaCFFs}, which can encourage changes in behaviour \cite{Caraban2019Nudging, Cox2016DesignFrictions}. 

Friction can be introduced into user interfaces in many ways, for example, by increasing temporal costs. Wang et al. \cite{Wang2014NudgeFacebookDelay} showed that time delays can encourage increased reflection before posting to social media. Lewis and Vogel \cite{Lewis2020Delays} used time delays to encourage users to transition to expert-mode menu selection techniques \cite{Lewis2020Delays}. When using an AI assistant, Bu\c{c}inca et al. \cite{BuccincaCFFs} required users to wait before displaying AI-generated suggestions, which encouraged more analytical thinking about the output and reduced over-reliance \cite{BuccincaCFFs}. Joshi and Vogel \cite{Joshi2025Ownership} proposed augmenting the prompt submission button so users had to press and hold it for a few seconds when submitting a shorter prompt. This effectively increased prompt length, but they did not verify the effect on psychological ownership.

\begin{figure*}[t!]
    \centering
    \includegraphics[width=0.6\textwidth]{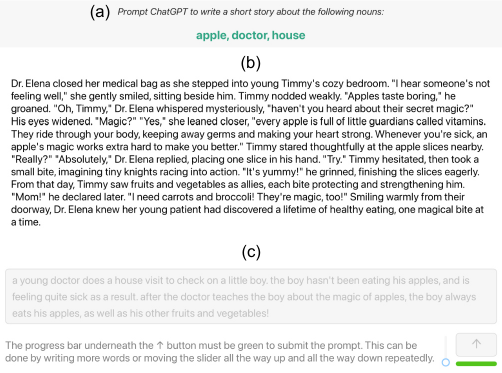}
    \caption{Experimental interface (shown with \f{motor}): (a) instructions and nouns the story had to be about; (b) the AI-generated short story; and (c) the augmented prompt entry interface.}
    \label{fig:interface}
\end{figure*}

Beyond increasing the temporal costs of submitting a shorter prompt, there are many other ways users could be encouraged to write longer prompts, and prior work has shown the effectiveness of other costs in non-AI contexts. For example, increasing the perception cost of reading text can reduce errors when transcribing it \cite{soboczenski2013increasing}, and increasing the perception and motor costs of viewing an interface (i.e., by ``brushing'' away ``frost'' on the screen) can improve spatial tasks \cite{cockburn2007FrostBrush}.

\medbreak

Overall, psychological ownership is important when writing with AI, and factors, including prompt length, can affect it. However, previous work has not evaluated the effect on psychological ownership for techniques designed to encourage writing longer prompts by introducing friction through costs.

%% file: 3-technique.tex
\section{Augmented Prompt Entry Techniques}

We evaluate the effectiveness of two different augmented prompt entry techniques for increasing prompt length and psychological ownership: one new technique that increases the motor cost of submitting a short prompt, and another from Joshi and Vogel that increases the temporal cost of submitting a short prompt \cite{Joshi2025Ownership}.\footnote{Both techniques are shown in detail in the accompanying video figure.} Both techniques are based on the idea of having to fill a `gauge' to submit a prompt; an idea that proved to be successful among several other ideas originally tested by Joshi and Vogel. We compare both of these to a baseline condition, where the user can press the prompt submission button normally, whenever they wish, and without any additional costs. 

\begin{table*}[h!]
    \centering
    \caption{Demographic information for Experiment 1, adapted from Masson et al. \cite{masson2023directgpt}.}
    \input{tables/exp1-demographics}
    \label{tab:exp1Demo}
\end{table*}

\subsection{Motor Cost}
Our new interaction technique introduces friction by increasing the \emph{motor cost} of submitting a short prompt. Below the prompt submission button is a small progress bar that represents a `gauge' that the user must fill. This can be done by writing more words in the prompt submission text box, or by `pumping' it full by moving a 45 pixel tall vertical slider up and down repeatedly (Figure \ref{fig:teaser}a). 

The progress bar width is the ratio of the number of words in the prompt to the \emph{word minimum}. We use 150 words to align with Joshi and Vogel's previous experiment \cite{Joshi2025Ownership}. 
When the user moves the slider up and down, the progress bar width increases. The number of movements the user must perform to fill the gauge is calculated by mapping the ratio of words typed to a \emph{movement maximum} (225 movements, which was determined through a pilot experiment; see Appendix \ref{sec:pilotExperiment} for more detail).
For example, if the user types 15 words, this means the gauge is 10\% full. To submit this prompt, the user must then move the slider up and down 203 times (90\% of 225). But if the user types 125 words, the gauge is 83\% full, so the user only has to move the slider 38 times. If the user types at least 150 words, they do not have to move the slider at all.

Moving the slider up and down to pump up the gauge does not have to be done continuously; the user could lift their mouse after a few movements and continue later. However, when they edit their prompt text, the gauge is reset to reflect the width from words typed. Any changes made by pumping or adding more words to the prompt trigger a 1 s `pulsing' animation to draw the user's attention to the progress being made. When the gauge is full, the progress bar turns green and the submit button becomes enabled for the user to press. Otherwise, the progress bar is blue and the submit button is disabled.

\subsection{Time Cost}
We also made minor adjustments to Joshi and Vogel's technique, which introduced a \emph{temporal cost} (cf. \cite{Joshi2025Ownership}; Figure \ref{fig:teaser}b). The technique also has a gauge containing a progress bar with a width that changes as the user types to represent a ratio of the \emph{word minimum}. But this time, if  the user's prompt length does not meet the word minimum, the user must press and hold the prompt submission button for a certain time. The time is a ratio of a \emph{wait maximum}, which is 60 s, as was done in their original paper. Unlike the motor cost, the user must press and hold in one act; all progress from pressing and holding is reset to zero when the mouse is released. The progress bar expands accordingly every 100 ms.  The same `pulsing' animation triggers when the user types or is holding the prompt submission button, and the progress bar turns from blue to green when full.

%% file: tables/exp1-demographics.tex
\small %
\begin{tabular}{lr|lr|lr|lr|lr}
\toprule
\multicolumn{2}{l|}{Gender} & \multicolumn{2}{l|}{Age} & \multicolumn{2}{l|}{Education} & \multicolumn{2}{l|}{Creative Writing Frequency} & \multicolumn{2}{l}{Prompt Engineering Familiarity}\\
\midrule
Men & 20 & 18-24 & 1 & High School & 3 & Weekly & 3 & Extremely & 2\\
Women & 9 & 25-34 & 5 & Some University (no credit) & 4 & Monthly & 8 & Moderately & 4\\
&& 35-44 & 16 & Technical Degree & 2 & Less than Monthly & 9 & Somewhat & 5 \\
&& 45-54 & 4 & Bachelor's Degree & 16 & Never & 9 & Slightly & 9\\
&& 55-64 & 1& Master's Degree  & 4 & &  & Not at All & 9\\
&& 65-74 & 2 &&&&
\end{tabular}

\begin{tabular}{lr|lr|lrlr|lrlr}
\\
\toprule
\multicolumn{2}{l|}{ChatGPT Frequency} & \multicolumn{2}{l|}{Weekly ChatGPT Use} & \multicolumn{4}{l|}{Other AI Services} & \multicolumn{4}{l}{AI Usage}\\
\midrule
Daily & 11 & 0 times & 3 & OpenAI Models & 20 & Dall-E & 3 & Write Emails & 10 & Get a Definition & 19\\
Weekly & 11 & 1-5 times & 8 & CoPilot & 4 & Midjourney & 2 & Write Papers/Essays & 4 & Explore a Topic & 20\\
Monthly & 5 & 5-10 times & 5 & Bing & 16 & Stable Diffusion & 1 & Write Stories & 9 & Brainstorming & 20\\
Less than Monthly & 2 & 10-30 times & 6 & Gemini & 19 & Other & 4 & Edit Text & 12 & Find References & 6\\
&& 30+ times & 7 & Grammarly & 9 &&& Generate Code & 2 & Clarification & 18 \\
&&&&&&&& Edit Code & 2 & Translate Text & 9\\
&&&&&&&& Debug Code & 1 & Other & 1\\
&&&&&&&& Generate Images & 20 &&\\
\end{tabular}

%% file: 4.1-exp-protocol.tex
\section{Experiment 1: Submission Techniques}
The goal of this experiment is to evaluate the effectiveness of the interaction techniques that introduce friction through costs for increasing prompt length and psychological ownership. This is a within-subjects experiment where  participants wrote prompts to generate short stories using each technique.

\subsection{Participants}
We recruited 35 participants from the online crowdsourcing platform, Prolific.\footnote{\href{https://www.prolific.com}{https://www.prolific.com}} Participants were restricted to the United States and Canada, and those who had successfully completed at least 2,500 tasks on the platform with a 99-100\% approval rating. We manually examined all open-ended responses to identify participants who experienced technical issues or who cheated by not making enough keystrokes in the prompt, suggesting they copied and pasted text into the prompt entry text box. Six (17\%) were removed from the analysis for these reasons, leaving 29 valid responses (Table \ref{tab:exp1Demo}). All self-reported English reading and writing proficiency (all $\geq$ 4 on a 7-point scale). Participants received \$7.50 after completing the experiment.

\subsection{Apparatus and Task}
The experiment interface was served as a Node.js and React web application. It resembled ChatGPT, with a few minor changes (Figure \ref{fig:interface}). The top of the interface displayed instructions and a ``triad'' of nouns to write about, which was adapted from Foley et al. \cite{Foley2020Composition} and previously used in Joshi and Vogel's experiment \cite{Joshi2025Ownership}. Writing short stories based on a series of nouns is useful for this type of experiment as it encourages impromptu creative thinking during experiments \cite{dunlop2016Images}, allows for increased experimental control, and also encourages high personal investment to write.
At the bottom, was the prompt entry interface. The prompt text box was 46 pixels tall by default and grew by 46 pixels as new lines were typed, until a maximum height of 300 pixels. Beside it was the prompt submission button, which was modified, depending on the condition. Underneath the text box was a message that explained how the current technique worked.
After writing their prompt and submitting it, the generated story appeared gradually by words or phrases on top of the text box, giving participants a chance to read it. The story was generated using the OpenAI GPT-4.5 Preview model. After, the user pressed a \emph{Continue} button at the bottom right corner to conclude the trial.

\begin{table*}[h!]
    \centering
    \caption{Omnibus and post hoc statistical tests for Experiment 1: (a) \m{Word Count}, (b) \m{Text Similarity}, (c) \m{Psychological Ownership}, (d) \m{Mental Demand}, (e) \m{Physical Demand},  (f) \m{Effort}, and (g) \m{Frustration}.}
    \input{tables/exp1-results}
    \label{tab:exp1Significance}
\end{table*}

\subsection{Procedure}
The task was restricted to laptops and desktops. Through the Prolific system, participants accessed a link to the web application. First, they entered basic demographic information and read instructions. Next, they wrote a prompt to generate a short story, and then answered questions about their experience. They repeated this for all conditions. Finally, they answered questions about their overall thoughts and preferences. The entire experiment lasted roughly 30 minutes.

\subsection{Design}
This is a within-subjects experiment with one primary independent variable, \f{condition} (levels: \f{baseline}, \f{time}, \f{motor}). We suspected possible order effects of \f{condition}, where participants may write more for \f{baseline} after experiencing \f{time} or \f{motor} first. As such, to get a more accurate measure of \f{baseline}, participants always prompted with this \f{condition} first. The order of \f{time} and \f{motor} was randomly assigned. There was only one trial per \f{condition}.     
The primary measures are subjective questions (all 0-100 interval data).\footnote{See the supplementary materials for question wording and experimental data.} \m{Psychological Ownership} represents how much psychological ownership participants felt towards their stories. Like Joshi and Vogel's experiment \cite{Joshi2025Ownership}, this was done by asking four questions \cite{nicholes2017measuring, Caspi2011Collaboration} about \m{Personal Ownership}, \m{Responsibility}, \m{Personal Connection}, and \m{Emotional Connection} and averaging these results to create a composite measure. The internal consistency reliability score was high ($\alpha$ = .94), so this composite measure was appropriate to use for analysis.
We also asked about workload-related factors from the NASA-TLX: \m{Mental Demand}, \m{Physical Demand}, \m{Temporal Demand}, \m{Performance}, \m{Effort}, and \m{Frustration}.
In addition, the following objective measures were collected from logs: \m{Word Count}, how long the prompt was when it was submitted; \m{Text Similarity}, the semantic text similarity between the final prompt and generated story, calculated using Google's Universal Sentence Encoder (0-1 range; 1 for identical texts) \cite{cer2018universalSentenceEncoder}; and the \m{Initial Word Count}, how long the prompt was when the participant first tried to press the button (\f{time}) or move the slider (\f{motor}).

%% file: tables/exp1-results.tex
\newcommand{\tabspacelg}{\vspace{0.8em}}
\newcommand{\tabspacesm}{\vspace{0.25em}}

\small
\begin{tabular}{llrr|llrr|llrr|llrr}
\toprule
\multicolumn{4}{l}{\tabspacesm (a) \m{Word Count}} & \multicolumn{4}{l}{\tabspacesm (b) \m{Text Similarity}} & \multicolumn{4}{l}{\tabspacesm (c) \m{Psychological Ownership}} & \multicolumn{4}{l}{\tabspacesm (d) \m{Mental Demand}}\\
\multicolumn{4}{l}{\tabspacesm \friedmanNEffect{2}{29}{16.72}{.001}{.29}} & \multicolumn{4}{l}{\tabspacesm \friedmanNEffect{2}{29}{7.52}{.05}{.13}} & \multicolumn{4}{l}{\tabspacesm \friedmanNEffect{2}{29}{9.95}{.01}{.13}} & \multicolumn{4}{l}{\tabspacesm \friedmanNEffect{2}{29}{15.21}{.001}{.26}}\\
\multicolumn{2}{l}{\textit{comparisons}} & \multicolumn{2}{l}{\textit{p-value}} & \multicolumn{2}{l}{\textit{comparisons}} & \multicolumn{2}{l}{\textit{p-value}} & \multicolumn{2}{l}{\textit{comparisons}} & \multicolumn{2}{l}{\textit{p-value}} & \multicolumn{2}{l}{\textit{comparisons}} & \multicolumn{2}{l}{\textit{p-value}} \\ 
\midrule
\f{baseline} & \f{time} & < .001 & *** & \f{baseline} & \f{time} & .03 & * & \f{baseline} & \f{time} & .04 & * & \f{baseline} & \f{time} & .006 & **\\
\f{baseline} & \f{motor} & < .001 & *** & \f{baseline} & \f{time} & .01 & * & \f{baseline} & \f{motor} & .04 & * & \f{baseline} & \f{motor} & < .001 & ***\\
\f{time} & \f{motor} & .37 & \textit{n.s.} & \f{time} & \f{motor} & .67 & \textit{n.s.} & \f{time} & \f{motor} & .47 & \textit{n.s.} & \f{time} & \f{motor} & .92 & \textit{n.s.}\\
\end{tabular}
\medbreak
\begin{tabular}{llrr|llrr|llrr}
\toprule
\multicolumn{4}{l}{\tabspacesm (e) \m{Physical Demand}} & \multicolumn{4}{l}{\tabspacesm (f) \m{Effort}} & \multicolumn{4}{l}{\tabspacesm (g) \m{Frustration}}\\
\multicolumn{4}{l}{\tabspacesm \friedmanNEffect{2}{29}{37.90}{.001}{.65}} & \multicolumn{4}{l}{\tabspacesm \friedmanNEffect{2}{29}{23.79}{.001}{.41}} & \multicolumn{4}{l}{\tabspacesm \friedmanNEffect{2}{29}{16.53}{.001}{.28}}\\
\multicolumn{2}{l}{\textit{comparisons}} & \multicolumn{2}{l}{\textit{p-value}} & \multicolumn{2}{l}{\textit{comparisons}} & \multicolumn{2}{l}{\textit{p-value}} & \multicolumn{2}{l}{\textit{comparisons}} & \multicolumn{2}{l}{\textit{p-value}} \\ 
\midrule
\f{baseline} & \f{time} & < .001 & *** & \f{baseline} & \f{time} & < .001 & *** & \f{baseline} & \f{time} & < .001 & *** \\
\f{baseline} & \f{motor} & < .001 & *** & \f{baseline} & \f{time} & < .001 & *** & \f{baseline} & \f{motor} & .01 & *\\
\f{time} & \f{motor} & .02 & * & \f{time} & \f{motor} & .19 & \textit{n.s.} & \f{time} & \f{motor} & .57 & \textit{n.s.} \\
\end{tabular}

%% file: 4.2-exp-results.tex
\subsection{Results}
In the analysis to follow, we use a Friedman test and Wilcoxon signed-rank tests with Holm's corrections for multiple comparisons, and Spearman's correlation coefficients. \emph{Detailed test results are shown in Table \ref{tab:exp1Significance}}. All open-ended responses were grouped into broad themes by the first author.

\begin{figure}[b]
	\centering
	\includegraphics[width=0.47\textwidth, height=3cm]{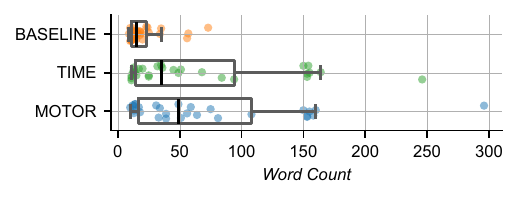}
	\caption{\m{Word Count} by \f{condition}.}
	\label{fig:wordCountNoFeedback} %
\end{figure}

\begin{figure}[b!]
	\centering
	\includegraphics[width=0.47\textwidth, height=3cm]{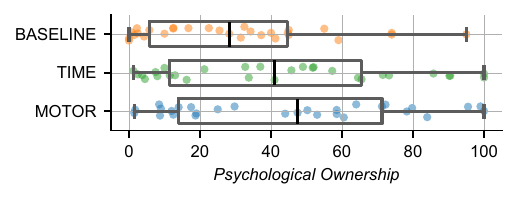}
	\caption{\m{Psychological Ownership} by \f{condition}. Higher scores correspond to higher feelings of psychological ownership.}
	\label{fig:ownershipNoFeedback} %
\end{figure}

\subsubsection{Word Count}
Overall, both interaction techniques led to significantly longer prompts (Figure \ref{fig:wordCountNoFeedback}). There was a significant effect of \f{condition} on \m{Word Count}, with post hoc tests revealing that \f{time} (\median{35}, \iqr{80}) and \f{motor} (\median{49}, \iqr{91}) both led to longer prompts than \f{baseline} (\median{15}, \iqr{23}). \m{Word Count} more than doubled for \f{time} and tripled for \f{motor}.
Seven participants (24\%) noted how the interaction techniques \ppquote{gave some time to reflect}{P14}, and encouraged them to \ppquote{write more and add more detail}{P22}.

\begin{figure*}[t!]
	\centering
	\includegraphics[width=\textwidth, height=3cm]{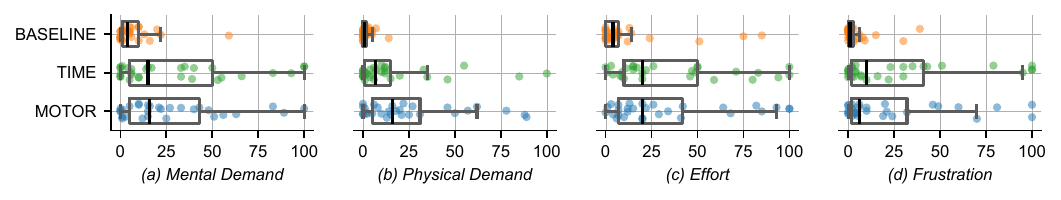}
	\caption{\m{Mental Demand}, \m{Physical Demand}, \m{Frustration}, and \m{Effort} by \f{condition}.}
	\label{fig:tlxNoFeedback} %
\end{figure*}

\subsubsection{Text Similarity}
Both interaction techniques resulted in higher similarities between the prompt and output. A significant effect of \f{condition} on \m{Text Similarity} and post hoc tests revealed that the generated stories for \f{time} (\median{.37}, \iqr{.18}) and \f{motor} (\median{.36}, \iqr{.12}) shared more similarities with the prompts than \f{baseline} (\median{.28}, \iqr{.15}). \m{Text Similarity} was moderately positively correlated with \m{Word Count} (\rs{.48}{.0001}), suggesting that as participants wrote longer prompts, the resulting stories more closely resembled them.

\subsubsection{Mechanics}
Although there were significant improvements in \m{Word Count} overall, \f{time} and \f{motor} did not work for all participants.
For \f{time}, participants tried to push the button 10 times on average. When comparing participants' \m{Initial Word Count} to their final \m{Word Count}, participants wrote significantly fewer words initially (\p{.01}; \median{20}, \iqr{36}). Twelve (41\%) increased their \m{Word Count} (\median{42.5}, \iqr{81.5} for differences), with 4 (14\%) increasing by 50 words or more. However, majority (16, 55\%) showed no differences between their \m{Initial Word Count} and final \m{Word Count}, and of these, only 1 wrote a prompt that was at least 150 words. These results are similar to Joshi and Vogel's original experiment \cite{Joshi2025Ownership}.

We observed a similar trend with \f{motor}. Participants tried to move the slider 6 times on average. Participants wrote significantly fewer words initially (\p{.01}; \median{31}, \iqr{28}). Twelve (41\%) increased their \m{Word Count} from their \m{Initial Word Count} (\median{31.5}, \iqr{52.25} for differences), with 5 (17\%) increasing by 50 words or more. Majority of participants (16, 55\%) had no difference, and of these, 5 (17\%) already wrote prompts that were at least 150 words long, meaning they did not have to move the slider at all.

\subsubsection{Psychological Ownership}
Both interaction techniques led to significantly higher feelings of psychological ownership (Figure \ref{fig:ownershipNoFeedback}). A significant effect of \f{condition} on \m{Psychological Ownership} with post hoc tests revealed that \f{time} (\median{41}, \iqr{54}) and \f{motor} (\median{47.5}, \iqr{57.25}) led to significantly higher \m{Psychological Ownership} than \f{baseline} (\median{28.25}, \iqr{39}), which was supported by comments like: \ppquote{I feel [like the time delay] was prompting me to be more involved in the story details. I would  spend more time on the prompt itself. It made the story a bit more personal}{P15}.

Although participants wrote longer prompts with \f{time} and \f{motor}, \m{Word Count} and \m{Text Similarity} were only weakly positively correlated with \m{Psychological Ownership} (both \rs{.28}{.01}). This may suggest that workload-related factors of the techniques contributed more to the higher feelings of psychological ownership than just prompt length and similarity.

\subsubsection{Task Workload}
Both interaction techniques led to significantly higher \m{Mental Demand}, \m{Physical Demand}, \m{Effort}, and \m{Frustration} (Figure \ref{fig:tlxNoFeedback}).
For \m{Mental Demand}, scores were lower for \f{baseline} (\median{4}, \iqr{9}) than \f{time} (\median{15}, \iqr{45}) and \f{motor} (\median{16}, \iqr{38}). \m{Mental Demand} was moderately positively correlated with \m{Word Count} (\rs{.48}{.0001}) and \m{Psychological Ownership}(\rs{.5}{.0001}).
\f{motor} had the highest \m{Physical Demand} (\median{16}, \iqr{26}), followed by \f{time} (\median{7}, \iqr{14}), and \f{baseline} (\median{1}, \iqr{2}).
For \m{Effort}, \f{baseline} (\median{4}, \iqr{7}) resulted in lower scores than \f{time} (\median{20}, \iqr{40}) and \f{motor} (\median{20}, \iqr{35}). \m{Effort} was weakly positively correlated with \m{Word Count} (\rs{.35}{.001}) and moderately positively correlated with \m{Psychological Ownership} (\rs{.42}{.0001}).
\f{baseline} (\median{1}, \iqr{3}) had lower \m{Frustration} than \f{time} (\median{10}, \iqr{39}) and \f{motor} (\median{6}, \iqr{30}).
No other metrics or correlations were significant.

These findings suggest that the increased \m{Mental Demand} and \m{Effort} required to submit prompts with \f{time} and \f{motor} contributed to higher feelings of \m{Psychological Ownership}. This was supported by comments like: \ppquote{I think [the slider] encourages me to add more of a personal input in the story that is submitted. Although it was a bit annoying, the end result felt more of a personal effort than not}{P20}; 
and \ppquote{having to move the slider handle up and down felt like work because it's something I had to do in order to submit the prompt. It made me feel a slightly deeper connection since I sort of had to work for it}{P18}.

\subsection{Summary}
Overall, our results show that increasing the temporal and motor cost of submitting a prompt can encourage users to write longer prompts. These techniques required additional mental and physical demand, and were more effortful and frustrating. However, the increased prompt length, text similarity, mental demand, and effort of these techniques all contributed to higher feelings of psychological ownership. These findings validate and significantly extend those of Joshi and Vogel's initial experiment \cite{Joshi2025Ownership} and provide evidence that incorporating motor costs into the prompt submission process can improve psychological ownership.

%% file: 5-exp2.tex
\begin{table*}[h!]
    \centering
    \caption{Demographic information for Experiment 2, adapted from Masson et al. \cite{masson2023directgpt}.}
    \input{tables/exp2-demographics}
    \label{tab:exp2Demo}
\end{table*}

\section{Receiving Suggestions from AI}
Encouraged by these results, we were interested in further augmenting these techniques to see if they could further improve psychological ownership.
Recall that prior work shows that receiving suggestions from AI can lower \cite{Lee2022CoAuthor, dhillon2024ScaffoldingOwnership} or increase \cite{Lehmann2022SuggestionList} psychological ownership depending on the interaction. Even in writing scenarios that do not involve AI, receiving feedback can help writers \cite{Kumaran2021FeedbackEarly} and suggestions can encourage ideation and creativity, and help to overcome writer's block \cite{TimedSuggestions2015}, which may occur when writing longer prompts \cite{Joshi2025Ownership}. Receiving suggestions while waiting or moving the slider may further encourage users to write longer prompts, which may further increase mental demand, effort, and psychological ownership.

We propose augmenting these techniques to display suggestions for how the prompt can be expanded when the user stops writing. After the user stops writing for 1 second, a blue tooltip appears on the top right corner of the prompt submission box (Figure \ref{fig:teaser}c). This resembles the ``on-idle'' condition from Siangliulue et al. \cite{TimedSuggestions2015}. The tooltip contains a suggestion, phrased as a question to encourage more thought \cite{AIQuestions}, that is one sentence long for how the prompt can be expanded or improved to incorporate more details (e.g., \emph{Consider: how does the relationship between the characters influence the story?}) These suggestions are generated using GPT 3.5 Turbo for increased speed.\footnote{See the supplementary materials for the prompt used to generate the suggestion.} Importantly, the suggestion is not automatically incorporated as edits to the prompt as this may reduce psychological ownership \cite{Lehmann2022SuggestionList}. The tooltip stays on the screen at maximum opacity for 4 seconds, then gradually fades away for 3 seconds, mimicking the mechanics of prior work on displaying notifications \cite{OppNudges}. This gives users time to choose whether they would like to incorporate these suggestions or not. If the user begins typing again, the tooltip does not show up (unless they again stop typing for 1 second). Otherwise, the tooltip reappears after 1 second.

\begin{table*}[h!]
    \centering
    \caption{Omnibus and post hoc statistical tests for Experiment 2: (a) \m{Word Count}, (b) \m{Text Similarity}, (c) \m{Psychological Ownership}, (d) \m{Mental Demand}, (e) \m{Physical Demand},  (f) \m{Effort}, (g) \m{Frustration}, and (h) \m{Number of Suggestions}.}
    \input{tables/exp2-results}
    \label{tab:exp2Significance}
\end{table*}

\begin{figure}[b!]
	\centering
	\includegraphics[width=0.47\textwidth, height=3cm]{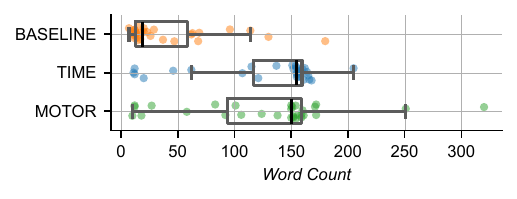}
	\caption{\m{Word Count} by \f{condition} when receiving AI-generated suggestions.}
	\label{fig:wordCountFeedback} %
\end{figure}

\section{Experiment 2: Prompt Suggestions}
We recruited 30 new participants using the same inclusion and exclusion criteria as before. No participants had to be filtered out (Table \ref{tab:exp2Demo}). Participants on Prolific are known for producing high-quality data \cite{CrowdsourcedDataQuality}, which could explain why no data had to be dropped.
All self-reported English reading and writing proficiency (all $\geq$ 4 on a 7-point scale). The task and procedure were both identical to that of Experiment 1.
The apparatus was mostly the same; however, each technique was further enhanced to display suggestions (Figure \ref{fig:teaser}c). These suggestions were integrated into the same interaction techniques and baseline as before. For all conditions, the prompt submission button was disabled until at least one suggestion was shown. 
The design was mostly identical to that of Experiment 1; however, we also considered \m{Number of Suggestions}, the number of suggestions each participant saw for a single \f{condition}.

\subsection{Results}
We use the same statistical tests as Experiment 1, with \emph{detailed statistical test results in Table \ref{tab:exp2Significance}}.

\subsubsection{Word Count}
Both interaction techniques led to significantly longer prompts (Figure \ref{fig:wordCountFeedback}). A significant effect of \f{condition} on \m{Word Count} and post hoc tests showed that \f{baseline} (\median{19}, \iqr{45.25}) resulted in shorter prompts than \f{time} (\median{155}, \iqr{43}) and \f{motor} (\median{150.5}, \iqr{64.75}). This represents a difference of over 100 words.

\subsubsection{Text Similarity}
Both interaction techniques also led to more text similarity. A significant effect of \m{Text Similarity} on \f{condition} revealed that prompts written with \f{baseline} (\median{.32}, \iqr{.14}) were less similar than \f{time} (\median{.41}, \iqr{.13}) and \f{motor} (\median{.42}, \iqr{.14}). \m{Text Similarity} was moderately positively correlated with \m{Word Count} (\rs{.55}{.0001}).

\begin{figure}[b!]
	\centering
	\includegraphics[width=0.47\textwidth, height=3cm]{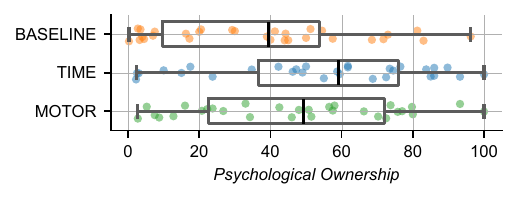}
	\caption{\m{Psychological Ownership} by \f{condition} when receiving AI-generated suggestions.}
	\label{fig:ownershipFeedback} %
\end{figure}

\subsubsection{Mechanics}
For \f{time}, participants pressed the button 7 times on average. When comparing the \m{Initial Word Count} to the final \m{Word Count}, participants wrote significantly fewer words initially (\p{.001}; \median{91.5}, \iqr{126.5}). Eleven (37\%) were consistent, of which 7 (23\%) consistently wrote at least 150 words. Majority of participants (18, 60\%) wrote more (\median{63}, \iqr{110}), with 11 (37\%) writing 50 words or more.

\begin{figure*}[h!]
	\centering
	\includegraphics[width=\textwidth, height=3cm]{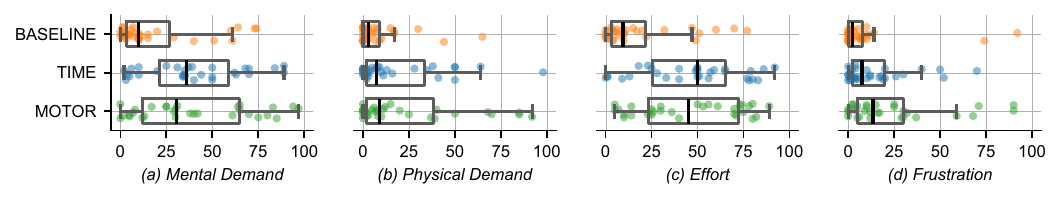}
	\caption{\m{Mental Demand}, \m{Physical Demand}, \m{Frustration}, and \m{Effort} by \f{condition} when receiving AI-generated suggestions.}
	\label{fig:tlxFeedback} %
\end{figure*}

For \f{motor}, participants moved the slider 6 times on average. Again, participants wrote significantly fewer words initially (\p{.01}; \median{96.5}, \iqr{134.75}). Majority (20, 67\%) were consistent, with 13 (43\%) writing at least 150 words and not having to interact with the slider at all. Ten (33\%) wrote more (\median{85}, \iqr{53.5} for differences), with 8 (27\%) writing at least 50 words.

\subsubsection{Psychological Ownership}
Both interaction techniques led to higher feelings of psychological ownership (Figure \ref{fig:ownershipFeedback}). A significant effect of \f{condition} on \m{Psychological Ownership} and post hoc tests showed that \f{time} (\median{59.12}, \iqr{39.38}) led to the highest \m{Psychological Ownership}, followed by \f{motor} (\median{49.38}, \iqr{49.31}), and \f{baseline} (\median{39.5}, \iqr{44}). \m{Psychological Ownership} was weakly positively correlated with \m{Word Count} (\rs{.37}{.001}) and \m{Text Similarity} (\rs{.32}{.01}).

\subsubsection{Task Workload}
Both interaction techniques increased \m{Mental Demand}, \m{Physical Demand}, \m{Effort}, and \m{Frustration} (Figure \ref{fig:tlxFeedback}).
For \m{Mental Demand}, \f{baseline} (\median{10}, \iqr{23.5}) resulted in lower scores than both \f{time} (\median{36}, \iqr{37.75}) and \f{motor} (\median{30.5}, \iqr{52.75}). \m{Mental Demand} was moderately positively correlated with \m{Word Count} and strongly positively correlated with \m{Psychological Ownership} (\rs{.62}{.0001}).
\f{baseline} (\median{3}, \iqr{8.75}) required less \m{Physical Demand} than \f{time} (\median{7.5}, \iqr{31.25}) and \f{motor} (\median{9}, \iqr{36.5}). There was a weak positive correlation between \m{Physical Demand} and \m{Psychological Ownership} (\rs{.35}{.001}).
For \m{Effort}, \f{baseline} (\median{9.5}, \iqr{18.25}) was less demanding than \f{time} (\median{50}, \iqr{39.25}) and \f{motor} (\median{45}, \iqr{49}). \m{Effort} was moderately positively correlated with \m{Word Count} (\rs{.43}{.0001}) and \m{Psychological Ownership} (\rs{.5}{.0001}).
\f{motor} (\median{13.5}, \iqr{25}) resulted in higher \m{Frustration} than \f{baseline} (\median{2.5}, \iqr{7.75}).
No other metrics or correlations were statistically significant.

\subsubsection{Number of Suggestions}
Participants received more suggestions when it was integrated into the two interaction techniques. A significant effect of \f{condition} on \m{Number of Suggestions} and post hoc tests revealed that participants received fewer suggestions with \f{baseline}(\median{4.5}, \iqr{8.25}) than \f{time}(\median{22}, \iqr{12}) and \f{motor} (\median{23}, \iqr{14.5}). \m{Number of Suggestions} was moderately positively correlated with \m{Word Count} (\rs{.55}{.0001}), \m{Text Similarity} (\rs{.45}{.0001}), \m{Mental Demand} (\rs{.49}{.0001}), and \m{Effort} (\rs{.54}{.0001}); and weakly positively correlated with \m{Physical Demand} (\rs{.22}{.05}), \m{Frustration} (\rs{.25}{.05}), and \m{Psychological Ownership} (\rs{.26}{.05}). 

Although four participants (13\%) noted that the suggestions were \ppquote{distracting or annoying}{P17}, 24 (80\%) appreciated them, as they helped when participants \ppquote{started to feel a bit stuck}{P10} and encouraged more thought, for example: \ppquote{[the suggestions] got me really thinking and making this more of a collaborative process than just entering a prompt and taking what the AI generated}{P28}.

%% file: tables/exp2-demographics.tex
\small %
\begin{tabular}{lr|lr|lr|lr|lr}
\toprule
\multicolumn{2}{l|}{Gender} & \multicolumn{2}{l|}{Age} & \multicolumn{2}{l|}{Education} & \multicolumn{2}{l|}{Creative Writing Frequency} & \multicolumn{2}{l}{Prompt Engineering Familiarity}\\
\midrule
Men & 15 & 18-24 & 2 & High School & 4 & Daily & 1 & Extremely & 3\\
Women & 15 & 25-34 & 6 & Some University (no credit) & 2 & Weekly & 6 & Moderately & 6\\
&& 35-44 & 6 & Technical Degree & 2 & Monthly & 6 & Somewhat & 5 \\
&& 45-54 & 10 & Bachelor's Degree & 17 & Less than Monthly & 12 & Slightly & 7\\
&& 55-64& 4& Master's Degree & 4 & Never & 5 & Not at All & 9\\
&& 65-74 & 2 & Doctorate & 1 &&
\end{tabular}

\begin{tabular}{lr|lr|lrlr|lrlr}
\\
\toprule
\multicolumn{2}{l|}{ChatGPT Frequency} & \multicolumn{2}{l|}{Weekly ChatGPT Use} & \multicolumn{4}{l|}{Other AI Services} & \multicolumn{4}{l}{AI Usage}\\
\midrule
Daily & 8 & 0 times & 4 & Nothing & 2 & Grammarly & 12 & Nothing & 2 & Generate Images & 15\\
Weekly & 12 & 1-5 times & 12 & OpenAI Models & 23 & Dall-E & 5 & Write Emails & 14 & Get a Definition & 17\\
Monthly & 5 & 5-10 times & 5 & CoPilot & 4 & Midjourney & 4 & Write Papers/Essays & 7 & Explore a Topic & 17\\
Less than Monthly & 3 & 10-30 times & 6 & Bing & 12 & Stable Diffusion & 3 & Write Stories & 9 & Brainstorming & 20\\
Never & 2 & 30+ times & 3 & Gemini & 18 & Other & 3 & Edit Text & 17 & Find References & 9 \\
&&&&&&&& Generate Code & 5 & Clarification & 13\\
&&&&&&&& Edit Code & 4 & Translate Text & 9\\
&&&&&&&& Debug Code & 3 & Other & 1\\
\end{tabular}

%% file: tables/exp2-results.tex
\newcommand{\tabspacelg}{\vspace{0.8em}}
\newcommand{\tabspacesm}{\vspace{0.25em}}

\small
\begin{tabular}{llrr|llrr|llrr|llrr}
\toprule
\multicolumn{4}{l}{\tabspacesm (a) \m{Word Count}} & \multicolumn{4}{l}{\tabspacesm (b) \m{Text Similarity}} & \multicolumn{4}{l}{\tabspacesm (c) \m{Psychological Ownership}} & \multicolumn{4}{l}{\tabspacesm (d) \m{Mental Demand}}\\
\multicolumn{4}{l}{\tabspacesm \friedmanNEffect{2}{30}{35.06}{.001}{.58}} & \multicolumn{4}{l}{\tabspacesm \friedmanNEffect{2}{30}{14.6}{.001}{.24}} & \multicolumn{4}{l}{\tabspacesm \friedmanNEffect{2}{30}{14.36}{.001}{.24}} & \multicolumn{4}{l}{\tabspacesm \friedmanNEffect{2}{30}{10.14}{.01}{.17}}\\
\multicolumn{2}{l}{\textit{comparisons}} & \multicolumn{2}{l}{\textit{p-value}} & \multicolumn{2}{l}{\textit{comparisons}} & \multicolumn{2}{l}{\textit{p-value}} & \multicolumn{2}{l}{\textit{comparisons}} & \multicolumn{2}{l}{\textit{p-value}} & \multicolumn{2}{l}{\textit{comparisons}} & \multicolumn{2}{l}{\textit{p-value}} \\ 
\midrule
\f{baseline} & \f{time} & < .001 & *** & \f{baseline} & \f{time} & < .001 & *** & \f{baseline} & \f{time} & < .001 & *** & \f{baseline} & \f{time} & .003 & **\\
\f{baseline} & \f{motor} & < .001 & *** & \f{baseline} & \f{time} & .002 & ** & \f{baseline} & \f{motor} & .005 & ** & \f{baseline} & \f{motor} & .007 & **\\
\f{time} & \f{motor} & .83 & \textit{n.s.} & \f{time} & \f{motor} & .79 & \textit{n.s.} & \f{time} & \f{motor} & .02 & * & \f{time} & \f{motor} & .98 & \textit{n.s.}\\
\end{tabular}
\medbreak
\begin{tabular}{llrr|llrr|llrr|llrr}
\toprule
\multicolumn{4}{l}{\tabspacesm (e) \m{Physical Demand}} & \multicolumn{4}{l}{\tabspacesm (f) \m{Effort}} & \multicolumn{4}{l}{\tabspacesm (g) \m{Frustration}} & \multicolumn{4}{l}{\tabspacesm (h) \m{Number of Suggestions}}\\
\multicolumn{4}{l}{\tabspacesm \friedmanNEffect{2}{30}{10.82}{.01}{.18}} & \multicolumn{4}{l}{\tabspacesm \friedmanNEffect{2}{30}{23.54}{.001}{.39}} & \multicolumn{4}{l}{\tabspacesm \friedmanNEffect{2}{30}{8.90}{.05}{.15}} & \multicolumn{4}{l}{\tabspacesm \friedmanNEffect{2}{30}{41.30}{.001}{.69}}\\
\multicolumn{2}{l}{\textit{comparisons}} & \multicolumn{2}{l}{\textit{p-value}} & \multicolumn{2}{l}{\textit{comparisons}} & \multicolumn{2}{l}{\textit{p-value}} & \multicolumn{2}{l}{\textit{comparisons}} & \multicolumn{2}{l}{\textit{p-value}} & \multicolumn{2}{l}{\textit{comparisons}} & \multicolumn{2}{l}{\textit{p-value}} \\ 
\midrule
\f{baseline} & \f{time} & .04 & * & \f{baseline} & \f{time} & < .001 & *** & \f{baseline} & \f{time} & .05 & \textit{n.s.} & \f{baseline} & \f{time} & < .001 & *** \\
\f{baseline} & \f{motor} & .007 & ** & \f{baseline} & \f{time} & < .001 & *** & \f{baseline} & \f{motor} & .003 & ** &  \f{baseline} & \f{motor} & < .001 & ***\\
\f{time} & \f{motor} & .14 & \textit{n.s.} & \f{time} & \f{motor} & .74 & \textit{n.s.} & \f{time} & \f{motor} & .05 & \textit{n.s.} & \f{time} & \f{motor} & .85 & \textit{n.s.} \\
\end{tabular}

%% file: 5.1-comparing.tex
\subsection{The Effects of Receiving Suggestions}
We use Mann-Whitney U tests to make between-subjects comparisons with Experiment 1 to examine the effect of receiving AI-generated suggestions (levels: \f{none}, from Experiment 1; and \f{suggestions}, from Experiment 2). \emph{Statistical test details are in Table \ref{tab:exp2Comparisons}.}

\begin{table*}[h!]
    \centering
    \caption{Between-subjects statistical tests comparing \f{none} and \f{suggestions} for: (a) \f{baseline}, (b) \f{time}, and (c) \f{motor}.}
    \input{tables/exp2-comparisons}
    \label{tab:exp2Comparisons}
\end{table*}

Overall, participants wrote over 100 more words with both interaction techniques when they received suggestions. For both \f{time} and \f{motor}, \m{Word Count} was significantly higher with \f{suggestions}, but receiving suggestions did not affect \f{baseline}. We observed a similar trend for \m{Text Similarity}, where \f{suggestions} increased similarity for \f{time} and \f{motor}, but not \f{baseline}. For \f{time}, participants made their first submission attempt when \m{Initial Word Count} was significantly longer with \f{suggestions}, but this did not occur for \f{motor}. This could explain why there were significant differences in \m{Psychological Ownership} between \f{time} and \f{motor} for \f{suggestions}.

Even though participants wrote significantly more words and had more similar texts with \f{time} and \f{motor}, this did not further improve \m{Psychological Ownership} as no comparisons with \f{none} were significant. For \f{baseline}, having \f{suggestions} led to significantly higher \m{Mental Demand}, \m{Physical Demand}, \m{Temporal Demand}, and \m{Effort}, but there were no differences for \f{time} and \f{motor}.

\subsection{Summary}
Overall, our results show that providing AI-generated suggestions can help users write even longer prompts when using techniques that increase the temporal and motor costs of submitting prompts. However, this did not further improve psychological ownership when compared to providing no suggestions. 

%% file: tables/exp2-comparisons.tex
\newcommand{\tabspacelg}{\vspace{0.8em}}
\newcommand{\tabspacesm}{\vspace{0.25em}}

\small
\begin{tabular}{lrr|rr|rr}
\toprule
& \multicolumn{2}{l}{\tabspacesm (a) \f{baseline}} & \multicolumn{2}{l}{\tabspacesm (b) \f{time}} & \multicolumn{2}{l}{\tabspacesm (c) \f{motor}}\\
\textit{metric} & \multicolumn{2}{l}{\textit{p-value}} & \multicolumn{2}{l}{\textit{p-value}} & \multicolumn{2}{l}{\textit{p-value}} \\ 
\midrule
\m{Word Count} & .10 & \textit{n.s.} & < .001 & *** & .003 & ** \\
\m{Text Similarity} & .22 & \textit{n.s.} & .04 & * & .02 & * \\
\m{Initial Word Count} &  & & .01 & * & .44 & \textit{n.s.}\\
\m{Psychological Ownership} & .25 & \textit{n.s.} & .16 & \textit{n.s.} & .73 & \textit{n.s.}\\
\m{Mental Demand} & .008 & ** & .10 & \textit{n.s.} & .11 & \textit{n.s.}\\
\m{Physical Demand} & .03 & * & .92 & \textit{n.s.} & .46 & \textit{n.s.} \\
\m{Temporal Demand} & .01 & * & .22 & \textit{n.s.} & .20 & \textit{n.s.} \\
\m{Effort} & .04 & * & .10 & \textit{n.s.} & .07 & \textit{n.s.}\\
\end{tabular}

%% file: 6-discussion.tex
\section{Discussion}
Our first experiment shows that introducing friction by increasing the time and motor costs of submitting short prompts can encourage users to write longer prompts. This typically required more mental demand and effort, but critically, it also increased psychological ownership. Our second experiment further enhanced these two techniques by incorporating AI-generated suggestions into the prompt entry interface. This further increased prompt length but did not further improve psychological ownership when compared to no suggestions.

These findings expand and provide additional nuance to Joshi and Vogel's original experiment \cite{Joshi2025Ownership}. Even though receiving suggestions made it easier for participants to write more and resulted in more text similarity, this did not lead to differences in psychological ownership. One possibility is that psychological ownership plateaued with longer prompts \cite{Joshi2025Ownership}, but another is that other factors beyond prompt length and text similarity were critical to increasing it. It could have been that workload-related factors were the most important contributors to increased psychological ownership. Unlike word count and text similarity, which were only weakly positively correlated with psychological ownership, mental demand and effort were notably both moderately positively correlated with psychological ownership in both experiments. To verify this, we conducted a variable importance analysis using the random forest method, and found that mental demand was the most important contributor to increased psychological ownership. Prior work also suggests that receiving suggestions from AI can affect what users think and write about \cite{OpiniatedAISuggestions}, so another possibility is that receiving suggestions affected how participants thought about their stories and felt less psychological ownership as a result.

This is not to say that encouraging users to write prompt length is not a valuable outcome; rather, encouraging users to write longer prompts can be a way to encourage more thought and improve psychological ownership.
As such, future work could continue to explore ways to increase prompt length and mental demand by imposing other costs when submitting short prompts, like perceptual costs, which have shown promise in non-AI contexts \cite{cockburn2007FrostBrush, soboczenski2013increasing}. 

We believe these ideas of increasing the costs of submitting shorter prompts could be valuable for other types of human-AI collaborations, like requiring users to submit detailed sketches when generating images to improve higher psychological ownership. It could also be useful in non-AI writing contexts. For example, incorporating such techniques into survey responses could encourage respondents to write longer open-ended responses. 

Encouraging users to write longer prompts is just one technique for increasing psychological ownership that can be easily integrated into current, chat-based interfaces. Future work should explore other mechanisms that encourage users to put more thought and effort into their interactions with generative AI writing assistants for increased psychological ownership.

\subsection{Limitations}
The baseline condition was always presented first to reduce potential order effects (writing more in \f{baseline} after experiencing \f{time} or \f{motor} first). However, we acknowledge that always experiencing the baseline condition first may have primed participants in ways that could have influenced their performance in subsequent conditions (e.g., by increasing task familiarity or becoming more inspired to write after completing the baseline trial).

Our results from the second experiment may also be limited by how we chose to display the prompt suggestions. Given the short writing task and the design of the two interaction techniques we tested, we felt that showing suggestions after 1 s of inactivity was appropriate; however, it may have been more appropriate to display them after a longer period or to display them on-demand (e.g., after the participant presses a button \cite{TimedSuggestions2015, StolenElephant}). On-demand suggestions may also result in increased psychological ownership due to the explicit action required from the user to receive a suggestion \cite{Lehmann2022SuggestionList}.

An ongoing challenge is integrating such techniques into generative AI writing tools. Users may not be willing engage with them for a real-world writing task that requires more urgency, but prior work suggests that additional interface augmentations could help, for example, by providing feedback and explanations for why these techniques exist \cite{nah2004study, egelman2010please}. In practice, these techniques may be better suited to being invoked at the user's request, aligning with ongoing trends to self-impose `lockout' mechanisms to manage technology use (e.g., \cite{LocknLol, KimLocknType}).

%% file: 7-conclusion.tex
\section{Conclusion}
We propose increasing the motor cost of submitting shorter prompts by requiring users to repeatedly move a slider up and down to fill a `gauge,' which can also be filled by writing more words in the prompt submission text box. Two experiments compared this approach to a technique from prior work that increases the temporal cost by requiring users to press and hold the prompt submission button and a baseline technique with no augmentations, with and without AI-generated suggestions. Both types of costs introduced friction to encourage users to write longer prompts and increased psychological ownership, but despite the even longer prompts written with AI-generated suggestions, this did not further improve psychological ownership.
Our results show that simple interaction techniques that introduce friction to increase the time and motor costs of submitting short prompts can encourage users to submit longer prompts with more details, which can increase mental demand and positively impact psychological ownership and human-AI collaborative writing.

%% file: 8-appendix.tex
\renewcommand\thefigure{\thesection.\arabic{figure}}
\renewcommand\thetable{\thesection.\arabic{table}}

\section{Pilot Experiment}
\label{sec:pilotExperiment}

We wanted to make sure that the two interaction techniques were similar in how long they took, as differences could have confounded our results. As the time cost technique required users to wait for 60 s in the extreme case, we wanted the extreme case of the motor cost technique to take roughly 60 s as well. To get a sense of how many slider movements would be needed to approximate 60 s, we asked 5 people within our research group to try moving the slider 100 times. The median duration for a single movement was 267 ms. Dividing 60,000 ms by 267 ms gives roughly 225 movements to approximate 60 s.